\documentclass[amssymb,prb,twocolumn,showpacs,superscriptaddress]{revtex4}
\usepackage{graphicx}
\usepackage{bm}
\begin{document}

\title{The ac magnetic response of mesoscopic type II superconductors} 
\author{Alexander D. Hern\'andez}
\affiliation{Laboratorio de
Superconductividad, Facultad de F\'{\i}sica-IMRE, Universidad de la
Habana, 10400, Ciudad Habana, Cuba.}
\affiliation{
Centro At\'{o}mico Bariloche and Instituto Balseiro, 8400 San Carlos  de
Bariloche, R\'{\i}o Negro, Argentina}
\author{Daniel Dom\'{\i}nguez}
\affiliation{
Centro At\'{o}mico Bariloche and Instituto Balseiro, 8400 San Carlos  de
Bariloche, R\'{\i}o Negro, Argentina}

\begin{abstract} 

The response of mesoscopic superconductors to an ac magnetic field is
numerically investigated on the basis of the time-dependent Ginzburg-Landau
equations (TDGL). We study the  dependence with frequency $\omega$ and 
dc magnetic field $H_{dc}$ of the linear 
ac susceptibility $\chi(H_{dc}, \omega)$ in square samples with dimensions of 
the order of the London penetration depth. 
At $H_{dc}=0$ the behavior of $\chi$ as a function of $\omega$ 
agrees very well with the two fluid model, and
the imaginary part of the ac susceptibility, $\chi"(\omega)$, 
shows a dissipative a maximum at 
the frequency $\nu_o=c^2/(4\pi \sigma\lambda^2)$. 
In the presence of a magnetic field a second dissipation maximum
appears at a frequency $\omega_p\ll\nu_0$. 
The most interesting behavior of 
mesoscopic superconductors can be observed 
in the $\chi(H_{dc})$ curves 
obtained at a fixed  frequency.
At a fixed number of vortices, $\chi"(H_{dc})$ continuously increases 
with increasing $H_{dc}$. We observe that the dissipation reaches
a maximum for magnetic fields right below the vortex penetration fields.
Then, after each vortex  penetration event, there is a sudden
suppression of the ac losses, showing discontinuities 
in $\chi"(H_{dc})$ at several values of $H_{dc}$. 
We show that these discontinuities are typical of the 
mesoscopic scale and disappear in macroscopic samples, which have a 
continuos behavior of $\chi(H_{dc})$. 
We argue that these discontinuities in $\chi(H_{dc})$ are
 due to the effect of {\it nascent vortices} 
which cause a large variation of the amplitude of the order parameter
near the surface before the entrance of vortices. 

\end{abstract}


\maketitle

\section{INTRODUCTION}

The response of superconductors to an ac magnetic field has been of 
interest for a long time,\cite{pippard,clem0,sridhar2} 
and particularly in the last years.
\cite{clem,brandt,geshk,beek,sonin,sridhar1,maeda1,%
mallozzi,portis,gomory,maeda2} 
The microwave surface impedance $Z_s=R_s-iX_s$ and the ac
magnetic susceptibility $\tilde\chi=\chi' + i\chi''$ 
have been extensively studied.
\cite{pippard,clem0,sridhar2,clem,brandt,geshk,beek,portis,sonin,sridhar1,maeda1,%
mallozzi,gomory,maeda2} 
Interesting 
behavior has been found for different values of the frequency, the magnetic 
field, the pinning force and the thermal
fluctuations,\cite{clem,brandt,geshk,beek}
whereas linear and nonlinear response appear depending on the strength of 
the ac signal.\cite{beek} 
The main interest so far has been on the electrodynamics of macroscopic 
samples containing a large number of vortices. 
In this case, it is possible to use 
phenomenological models to describe the macroscopic behavior 
of the superconducting samples.\cite{clem,brandt,geshk,beek}

Recently there has been an increasing interest in the study of vortex 
physics on a mesoscopic scale, 
\cite{buisson,bolech,bolle,geim98,geim00,peeters97,peeters98a,peeters99b,%
peeters00,peeters99c,peeters01,palacios,bruyndoncx,chibotaru,HD} 
a regime in which a small number of vortices are confined in a small sample. 
The behavior of 
mesoscopic superconductors is different from the behavior of 
bulk samples. In mesoscopic superconductors surface effects\cite{bean} 
are very important since the interaction of vortices with the surface 
currents is large.\cite{HD}
The magnetic properties strongly depend on the sample sizes and
geometry.\cite{peeters99b,peeters00,HD} 
The most studied geometries are the 
the mesoscopic disk \cite{buisson,geim98,geim00,peeters97,peeters98a,%
peeters99b,peeters00,peeters99c,peeters01,palacios} 
the mesoscopic slab\cite{bolech,bolle} and the mesoscopic square. 
\cite{bruyndoncx,chibotaru,HD}
In particular the mesoscopic samples can develop Abrikosov multivortex 
states \cite{peeters98a} and depending on the size of the sample 
it is possible to observe first or second order  transitions.\cite{peeters97} 
The electric charge of vortices, \cite{peeters01} a paramagnetic  
effect \cite{geim98,palacios} and the surface barrier
\cite{HD,bean} 
on a mesoscopic scale have also been studied. One interesting
characteristic of the magnetic properties of mesoscopic superconductors is the
behavior of the dc magnetization curves. In a mesoscopic scale, 
the vortices that are inside the sample induce a reinforcement of the surface 
barrier at fields greater than the first penetration field $H_{p}$.\cite{HD}
These new barriers allow for the existence of metastable states of constant 
vorticity as a function of magnetic field. 
Each metastable state becomes unstable at the $i$-th penetration field $H_{pi}$ 
in which vortices enter the sample and the magnetization has a discontinuous 
jump. These jumps have been observed numerically \cite{bolech,peeters97,HD} 
and experimentally in mesoscopic Al disks.\cite{peeters97} 

In the last years numerical simulations of the time dependent Ginzburg Landau 
(TDGL) equations have been an important tool in the study of the 
static and dynamic properties of superconductors.
\cite{bolech,frah,liu,kato,machida,aranson1,aranson2,aranson3,enomoto,%
alvarez,gropp} 
The TDGL model has been used to study the flux growth dynamics \cite{frah,liu}, 
the magnetic response\cite{bolech,kato} and the I-V characteristics of 
superconducting samples.\cite{machida} 
In particular, Enomoto {\it et al.}\cite{enomoto} 
using the TDGL equations studied 
the temperature dependence of the ac susceptibility at a fixed frequency 
in the absence of a dc magnetic field for a large sample. 
For increasing temperature they found that $\chi'$
exhibits a step-like change from a negative value ($\chi'=-1/(4 \pi)$ to zero, 
while $\chi"$ initially rises from zero, goes through a maximum, and then 
returns to a small value near $T_c$, which is 
qualitatively consistent with the overall behavior observed in experiments.

In this paper we perform numerical simulations of the time dependent
Ginzburg-Landau equation to study the ac magnetic response of type II
superconductors on a mesoscopic scale. 
We study the frequency and $H_{dc}$ dependence of the linear 
ac susceptibility $\tilde\chi(H_{dc}, \omega)=
\chi'(H_{dc}, \omega)+i\chi''(H_{dc}, \omega)$ of square samples 
with dimensions of 
the order of the London penetration depth $\lambda_L$. 
Our resuls were obtained in
the absence of bulk pinning and in the linear regime.

The paper is organized as follows. 
In  Sec.II we review the known results for the ac magnetic response of 
macroscopic superconductors, in order to later compare them with
our results in mesoscopic samples.
In particular, the two fluid model for the Meissner state 
and the Coffey and Clem model for the mixed state are reviewed.
In Sec.III we describe the TDGL equations used in the simulations.
In Sec.IV we present our numerical results for the ac magnetic response
of mesoscopic square samples. 
Sec.IV.A is devoted to the study of the frequency 
dependence of $\chi' $ and $\chi "$. 
We find that for $H_{dc}=0$ the behavior of 
the sample is well described by the two fluid model. 
At $\omega = \nu_{o}=c^2/(4\pi \sigma\lambda^2)$ 
a maximum in the imaginary part of 
$\tilde\chi(\omega,0)$ appears. 
In the Meissner state for $H_{dc} \sim H_p$, another maximum appears in 
$\chi"(\omega)$. These behavior, characterized by the presence of two 
dissipation maxima is also observed in the mixed state. 
At the same time, for increasing frequency  
$\chi'(\omega, H_{dc})$ changes from London screening to a perfect
diamagnetic state at  high frequency. This transition is due to a decrease
of the ac penetration length. 
In Sec.IV.B we study the $H_{dc}$ 
dependence of $\chi '$ and $\chi "$ at a fixed frequency.
Depending on the sample size, we find well defined mesoscopic and 
macroscopic behaviors. 
For mesoscopic samples we observe that, 
in the magnetic field ranges in which the  number of vortices is
constant,  $\chi"(H_{dc})$ continuously increases for increasing $H_{dc}$.
At the $i$-th vortex penetration field $H_{pi}$, 
the entrance of vortices produces a 
considerable suppression of the ac losses and 
$\chi"(H_{dc})$ decreases with a discontinuous  jump. 
On the contrary, samples of
macroscopic sizes show a continuous behavior in $\tilde\chi(H_{dc})$,
with $\chi"(H_{dc})$ monotonically increasing with $H_{dc}$. 
In Sec. V we study the time evolution of the order parameter in mesoscopic
superconductors. We show that at high frequencies and in mesoscopic 
samples,  the  vortices are fixed and only play a secondary role in the ac 
response of  the sample. We show that the
main dissipation mechanism in mesoscopic
superconductors is due to the effect of ``nascent vortices''.
They cause  large variations of the amplitude of the order parameter
at the boundary of the sample 
before the  entrance of vortices.
Finally in Sec.VI we give a summary of our results and conclusions.

\section{THE AC RESPONSE OF MACROSCOPIC SAMPLES: REVIEW OF KNOWN RESULTS}

\subsection{Two fluid model and  Coffey and Clem model}

When an ac magnetic field is applied $H(t)=H_{dc}+h_{ac}\cos(\omega t)$,
an effective complex penetration depth $\tilde{\lambda}(\omega)$ can be defined
assuming that the fluctuating field inside a semiinfinite sample for $x>0$ has
the form
\begin{equation}
\delta H = h_{ac}e^{-x/\tilde{\lambda}(\omega)}e^{-i\omega t}
\end{equation}
or from a generalization of London's expression for the current,
\begin{equation}
 \nabla\times{\mathbf J}(\omega) =
-\frac{c}{4\pi\tilde{\lambda}^2(\omega)}{\mathbf B}(\omega)\;.
\label{genLon}
\end{equation}
It can be related to the frequency dependent conductivity as
$\tilde{\lambda}(\omega)=\sqrt{\frac{ic^2}{4\pi\omega\tilde\sigma(\omega)}}$.
In the case of a normal metal with conductivity $\sigma_n$, 
the complex penetration depth is directly related to the skin depth 
$\delta_n=\sqrt{\frac{c^2}{2\pi\omega\sigma_n}}$
as $\tilde\lambda_n = (1+i)\delta_n/2$.
The complex permeability $\tilde\mu(\omega)$, or the complex susceptibility 
$\tilde\chi(\omega)=\chi'(\omega)+i\chi''(\omega)=
[\tilde\mu(\omega) -1]/4\pi$,
depend both on  $\tilde{\lambda}(\omega)$ and also on the specific shape of the
sample, since the ac penetration of magnetic fields is a surface
phenomenon. 
In the case of a square sample of size $L\times L$ 
an approximate expression for the permeability 
in the limit $\tilde\lambda \ll L$ is
\begin{equation}
\tilde \mu_{{square}} \approx
\frac{4\tilde\lambda}{L}\left(1-\frac{\tilde\lambda}{L}\right).
\label{musq}
\end{equation}
Exact expressions for $\tilde\mu$ for squares, cylinders and slabs 
can be found for example in Ref.\onlinecite{clem}.  
Sometimes the complex surface impedance $Z_s$ is used, 
it can be defined when $\tilde\lambda \ll L$ as
$Z_s=R_s - i X_s = i\omega\frac{4\pi}{c^2}\tilde\lambda(\omega)\;.$

In the Meissner state the ac response has been usually described 
with the two fluid model.\cite{twofluid,gittleman}
In the mixed state, the magnetic flux enters in the 
form of quantized vortices.
Therefore the dynamics of the vortices has to be included
in a description of the magnetic ac response of the mixed state.
\cite{clem,brandt,geshk,beek} 
In particular, Coffey and Clem have extended
the two fluid model by including the equation of motion
of vortices for small displacements. \cite{clem}  
Similar results were  obtained by
Brandt\cite{brandt} and by van der Beek {\it et al.}\cite{beek}
All these models assume that there is a large number
of vortices in the sample and a continuous description 
of the vortex lattice is used (valid for $B\gg H_{c1}$).

The two fluid model consists in writing the total current as the sum
of the supercurrent and the normal current, 
\begin{equation}
{\mathbf J}={\mathbf J}_s + {\mathbf J}_n
\label{jsjn}
\end{equation}
with 
\begin{equation}
{\mathbf J}_n = \sigma_n {\mathbf E} = -\frac{\sigma_n}{c}\frac{\partial {\mathbf
A}}{\partial t}
\label{jn}
\end{equation}
and the supercurrent given by the London model
\begin{equation}
\nabla \times {\mathbf J}_s = -\frac{c}{4\pi\lambda_L^2} ({\mathbf B}-{\mathbf
b}_v)\;.
\label{vorLon}
\end{equation}
Here $\lambda_L$ is the static London penetration depth and 
${\mathbf b}_v$ is the local vortex magnetic field.
In the absence of vortices, ${\mathbf b}_v=0$, 
the two fluid model has the
characteristic time $t_0=4\pi\lambda_L^2\sigma_n/c^2$ 
for the transformation of supercurrents into normal currents, since
${\mathbf J}_n = 
t_0{\partial{\mathbf J}_s}/{\partial t}$.
The Coffey and Clem model includes the equation of motion of vortices
for small  displacements ${\mathbf u}({\mathbf x},t)$
 from their equilibrium positions, 
\begin{equation}
\eta_v\dot{\mathbf u}+\kappa_p{\mathbf u} = \frac{1}{c}{\mathbf
J}\times\Phi_0\hat{\bm \alpha}
\label{vormot}
\end{equation}
where $\eta_v$ is the viscous drag coefficient, $\kappa_p$ is the 
restoring force constant (Labusch parameter) of a pinning potential
well and $\hat{\bm \alpha}$ the local
vortex direction. Here we have neglected thermal fluctuation effects.   
The local vortex magnetic field 
depends on the  vortex displacements  as $
{\mathbf b}_v=\nabla\times({\mathbf u}\times{\mathbf B})$.\cite{beek}

In the case of fields ${\mathbf H}\parallel{\mathbf h}_{ac}$ and parallel to the
surface, using Eqs.(\ref{genLon}),(\ref{jsjn})-%
(\ref{vormot})
the effective complex penetration depth\cite{clem}
can be obtained as 
\begin{equation}
\tilde\lambda^2=\lambda_L^2\frac{1+\left(\frac{\lambda_L^2}{\lambda_C^2}-i\omega
t_{ff}\right)^{-1}}{1-i\omega t_0}
=\frac{\lambda_L^2+\left(\frac{1}{\lambda_C^2}-i\frac{2}{\delta_{ff}^2}\right)^{-1}}
{1-i\frac{2\lambda_L^2}{\delta_n^2}}
\label{cofclem}
\end{equation}
with the Campbell penetrationd depth $\lambda_C^2=B\Phi_0/4\pi\kappa_p$,
the flux-flow time scale $t_{ff}=4\pi\lambda_L^2\sigma_{ff}/c^2$, the flux-flow conductivity
$\sigma_{ff}=c^2\eta_v/B\Phi_0$ and the flux-flow skin depth
$\delta_{ff}^2=c^2/2\pi\omega\sigma_{ff}$.
Since $\sigma_{ff} \sim \sigma_n H_{c2}/B$, we have $\sigma_{ff} > \sigma_n$
and $t_{ff} > t_0$. Therefore the highest characteristic frequency
is $\nu_0=1/t_0$, which for conventional superconductors is 
about $\nu_0\sim 10-100$ GHz.\cite{tscales}

\subsection{Meissner state}

In the absence of vortices the complex penetration depth is
simply $\tilde\lambda^2=\lambda_L^2/(1-i\omega
t_0)$.
This leads to a dissipation peak in $\chi''$ when $\omega t_0 \approx
1$. 
For large frequencies, $\omega t_0\gg 1$ (i.e.
$\delta_n\ll\lambda_L$), the system behaves as a normal metal 
(normal state skin depth effect) with: 
\begin{equation}
\lambda'=\lambda''\approx \lambda_L\sqrt{\frac{1}{2\omega
t_0}}
\label{Mlarge}
\end{equation}
Therefore dissipation goes as $\chi''\propto\lambda''/L\sim\omega^{-1/2}$ 
for large $\omega$.
For low frequencies,   $\omega t_0\ll 1$ (i.e.
$\delta_n\gg\lambda_L$), the system is dominated by the Meissner effect
and
\begin{equation}
\lambda' \approx \lambda_L \;,
\;\;\;\;\;\;\;\;\lambda''\approx\lambda_L\frac{1}{2}\omega t_0
\label{Msmall}
\end{equation}
and therefore dissipation goes as $\chi''\propto\lambda''/L\sim\omega$ 
for low $\omega$.

\subsection{Vortices without pinning}

In the presence of vortices, the interesting frequency range
is $\omega t_0 \ll 1$ since
for frequencies $\omega t_0 \gg 1$ the system
always behaves as in Eq.(\ref{Mlarge}), corresponding to
the normal state skin depth effect. 

Let us first discuss the case when there is no bulk-pinning,
$\kappa_p=0$ ($\lambda_C=\infty$).
For frequencies such that $\omega t_{ff} \gg 1$, the
system behaves similarly to the Meissner state, with a dissipation
peak at $\omega t_0 \approx 1$, as described previously.
For frequencies such that $\omega t_{ff} \ll 1$ the system is
dominated by the ``flux flow skin depth effect'' 
due to the flux flow conductivity
$\sigma_{ff}$.  In this limit $\tilde\lambda^2\approx
i\delta_{ff}^2/2=i\lambda_L^2/\omega t_{ff}$, and therefore the susceptibility
for low frequencies should diverge as $\tilde\chi \sim \omega ^{-1/2}$. This
divergence is cut-off by the finite system size. 
At low frequencies the real part of the effective penetration depth
 $\lambda'$ saturates to the system size $L$.
This leads to a dissipation peak  at a frequency\cite{geshk}
\begin{equation}
\omega_L \propto \frac{c^2}{L^2\sigma_{ff}} \propto \frac{B}{L^2}\;.
\label{peakL}
\end{equation}
And  for very low  frequencies such that $\omega \ll \omega_L$, 
dissipation diminishes linearly as
\begin{equation}
\chi'' \propto \frac{\sigma_{ff} L^2}{c^2}\omega.
\label{Lsmall}
\end{equation}

\subsection{Vortices with pinning}

In the presence of bulk pinning, $\kappa_p\not=0$, 
the relevant time scale is $t_C=t_{ff}\lambda_C^2/\lambda_L^2=\eta_v/\kappa_p$.
Now instead of the finite size peak at $\omega_L$, 
there is a vortex dissipation peak when $\omega t_C\approx 1$.
It is worth mentioning that this peak frequency,
\begin{equation} 
\omega_C \propto \frac{\kappa_p}{\eta_v}, 
\label{peakC}
\end{equation}
is independent of magnetic field. 	
For low frequencies, $\omega t_C \ll 1$, the real part of the
penetration depth tends to $\lambda'\approx\tilde\lambda_C
=\sqrt{\lambda_C^2+\lambda_L^2}$,
while the imaginary part $\lambda''\approx \omega t_C
\lambda_C^2/2\tilde\lambda_C$. Thus for low frequencies the
dissipation diminishes linearly with $\omega$ as
\begin{equation}
\chi''\approx \frac{1}{2\pi} \frac{\lambda_C^2}{\tilde\lambda_C
L}\omega t_C.
\label{Csmall}
\end{equation}
Therefore, in the presence of vortices there is a new dissipation peak,
in addition to the ``two fluid peak'',
either at $\omega_L$ or at $\omega_C$ depending on the importance of
pinning. At low frequencies the relevant length scale for penetration of
the ac field is either $L$ (in the absence of pinning)  or 
$\lambda_C$ (if pinning is important)
and the dissipation is linear with frequency in both cases.

\subsection{Surface barrier effects}

At the surface of a type II superconductor there is a potential barrier
that prevents the entrance (and the exit) of vortices. This barrier
has been first calculated by Bean and Livingston.\cite{bean}
The surface barrier inhibits the penetration of flux at $H_{c1}$, where
penetration is thermodinamically favorable. Instead of $H_{c1}$, 
vortices start to enter at the first penetration field $H_p > H_{c1}$.  
Above $H_p$, the Bean-Livinngston surface barrier  can stabilize metastable 
states with a non-equilibrium number of vortices.\cite{clemlt13,ternovskii} 
The interval of external magnetic
fields where these metastable states exist in macroscopic
superconductors has been 
obtained in Ref.~\onlinecite{clemlt13,ternovskii,burlachkov}.
It has been found that for fields $H>H_p$ the vortex array is separated
from the boundary surface by a vortex free region because of the effect of 
the surface barrier.
The length of this vortex free region\cite{clemlt13,ternovskii,burlachkov} is
$d_{SB}=\lambda_L\cosh^{-1}(H/B)$.
The ac magnetic response has been calculated by 
Sonin and Traito\cite{sonin} in this case. 
They assume that the number of vortices is fixed, and
they allow  the size of the vortex free region $d_{SB}$
to fluctuate with the rf field $h_{ac}e^{-i\omega t}$.
A  new dissipation peak is found at a frequency
\begin{equation} 
\omega_{SB}\approx \frac{\tanh^2(d_{SB}/\lambda_L)}{t_{ff}}\;,
\label{peakS}
\end{equation}
which for $B\ll H_p$ is $\omega_{SB}\approx 1/t_{ff}$. 
For low frequencies $\omega\ll\omega_{SB}$, the effective penetration
depth goes as 
\begin{eqnarray}
\lambda'&\approx&\lambda_L\coth(\frac{d_{SB}}{\lambda_L})\label{Ssmall}\\
\lambda''&\approx&\frac{\lambda_L^2}{\sinh^2(\frac{d_{SB}}{\lambda_L})}
\sqrt{\frac{2\pi\sigma_{ff}\omega}{c^2}}\nonumber\;.
\end{eqnarray}
Therefore, according to this approach, the dissipation should
go as $\chi''\propto\lambda''/L\sim\omega^{1/2}$ if 
surface barrier effects are important.
Experimentally, dissipation maxima attributed to surface barriers have
been measured, for example in Ref.\onlinecite{fabrega}.
In platelet samples with a perpendicular magnetic field, geometrical
effects enhance the surface barrier, giving place to a ``geometrical
barrier''. Morozov {\it et al.}\cite{morozov} have measured the effect of 
geometrical barriers in the ac response of a superconductor. 
In this case there is a
dissipation maximum as a function of magnetic field for $H\agt H_p$,
which is independent of frequency.

\section{MODEL AND DYNAMICS}

Our numerical simulations are carried out using the 
time-dependent Ginzburg-Landau equations complemented with the 
appropriate Maxwell equations. In the zero-electric potential 
gauge we have: \cite{kato,gropp} 
\begin{eqnarray} 
\frac{\partial \Psi}{\partial t} = \frac{1}{\eta} [(\nabla -
i{\mathbf A})^2 \Psi +(1-T)(1-|\Psi |^2)\Psi ]   \label{tdgl1}\\
\frac{\partial {\mathbf A}}{\partial t} = (1-T)\mbox{Im}[\Psi^* (\nabla -
i{\mathbf A})\Psi] -\kappa^2\nabla \times \nabla \times {\mathbf A}
\label{tdgl2}
\end{eqnarray}
where $\Psi$ and ${\mathbf A}$ are the order parameter and vector potential
respectively and  $T$ is the temperature. 
Equations (\ref{tdgl1}) and (\ref{tdgl2}) are in their dimensionless form.
Lengths have been scaled in units of 
the coherence length $\xi(0)$, times in units of
$t_0=4\pi \sigma_n\lambda_L^2/c^2$, ${\mathbf A}$ in units of $H_{c2}(0) \xi (0)$ and
temperatures in units of $T_c$. $\eta$ is equal to the ratio
of the characteristic time $t_0$ for the relaxation of $A$ and
the time $t_{GL}$ for the relaxation of $\Psi$:
$\eta=t_{GL}/t_0=c^2/(4\pi\sigma_n\kappa^2D)$, 
with $t_{GL}=\xi^2/D$, 
where $\sigma_n$ is the quasiparticle conductivity and $D$ is the
electron diffusion constant. For superconductors with magnetic impurities
we have $D_{imp}=c^2/(48\pi\kappa^2\sigma_n)$, 
and therefore $\eta=12$ in this case.

We solve the TDGL equations following the same procedure as
in our previous work in Ref.\onlinecite{HD},
using a standard finite difference 
discretization scheme.\cite{gropp} 
We discretize space in a rectangular mesh with spacings
$\Delta x$ and $\Delta y$ in each direction.
The order parameter and vector potentials are defined
at the nodes of the rectangular mesh ($\vec{r}=(I,J)$), and the 
link variables $U_{_\mu I,J}=\exp(-\imath \kappa h_{\mu} A_{\mu I,J})
\;\;\; (\mu=x,y)$ are introduced in order to maintain the gauge 
invariance under discretization.
We assume that the sample  has a square shape in the 
$x,y$ direction and it is infinite in the $z$ direction. 
We apply the magnetic field parallel to the $z$ direction, therefore 
the problem is reduced to two dimensions neglecting all derivatives along $z$. 
The symmetry of 
the problem implies for all mesh points $A_{I,J}=(A_{xI,J},A_{yI,J},0)$ and 
 ${\mathbf B}_{I,J}=(0,0,B_{zI,J})$, where 
 $B_{_zI,J}=(\nabla \times \vec{A})_z=(\partial_xA_{_yI,J}-\partial_yA_{_xI,J})$.

The dynamical equations must be complemented with the appropriate boundary
conditions for both the order parameter and the vector potential. 
For the order parameter we have used the boundary condition:
\begin{equation} 
(\Pi \Psi)^\perp =(\nabla -i\vec{A})^\perp \Psi =0 
\end{equation}
usually known as the superconductor-insulator 
boundary condition (S-I) because it implies that the perpendicular
component of the superconducting current is equal to zero at the 
surface ($\vec{J}_s^\perp=0$). This boundary condition also minimizes the 
free energy at the sample surface.

The ac magnetic field is introduced in the simulation through the 
boundary condition for the vector potential $A_{\mu I,J}$. We consider the
case where the ac magnetic field ${\mathbf h}_{ac}$ is parallel to the dc-component 
${\mathbf H}_{dc}$, and both fields are in the $z$ direction: 
$$B_z|_S=(\nabla \times {\mathbf A})_z|_S=H_{dc} + h_{ac}\cos(\omega t)\;;$$
this expression is evaluated at the sample surface.
We study the response of a superconductor in the linear regime, i.e.
 $h_{ac} \ll H_{dc}$.
The time dependence of the sample magnetization can be obtained from the
magnetic induction averaged over the sample $\langle B(t) \rangle$ 
through the relation:
$$4 \pi M(t)=\langle B(t) \rangle - H(t)$$
The ac magnetic susceptibilities are obtained from the Fourier transform of 
$M(t)$:
\begin{eqnarray} 
\chi '(H_{dc},\omega ) = \frac{1}{\pi h_{ac}} \int_0^{2 \pi} M(t)\cos(\omega t)
d(\omega t) \nonumber \\
\chi "(H_{dc},\omega ) = \frac{1}{\pi h_{ac}} \int_0^{2 \pi} M(t)\sin(\omega t)
d(\omega t)\;.
\end{eqnarray}

In what follows, we have solved the TDGL equations numerically 
for a type II superconductor with $\kappa=2$, 
mean field temperature $T=0.5$ and the parameter $\eta=12$. 
We used a spatial discretization of $\Delta x= \Delta y=0.5 \xi$ 
and, in order to make efficient calculations, we 
have chosen adequately the time step
with values  $\Delta t \le 0.015 t_0$.

\section{THE AC SUSCEPTIBILITY IN MESOSCOPIC SUPERCONDUCTORS}

In a mesoscopic sample the behavior  of the ac magnetic response in the
presence of a dc magnetic field can be different than in the macroscopic
case. On one hand,
the models typically used for macroscopic samples (see Sec.II),
assume a high density of vortices (giving an almost uniform magnetic field
profile inside the sample\cite{brandt}), applied magnetic fields 
such that $H_{c1}\ll H \ll H_{c2}$, and very large (semiinfinite) samples.
On the other hand, in mesoscopic samples there is a 
small number of vortices, surface barriers can not be neglected and
the finite size of the sample is important.
Here we perform simulations of squares samples of sizes    
from $5 \lambda \times 5 \lambda$ to $20 \lambda \times 20 \lambda$ 
 using the TDGL equations.

\begin{figure}[htb]
\centerline{\includegraphics[width=8.5cm]{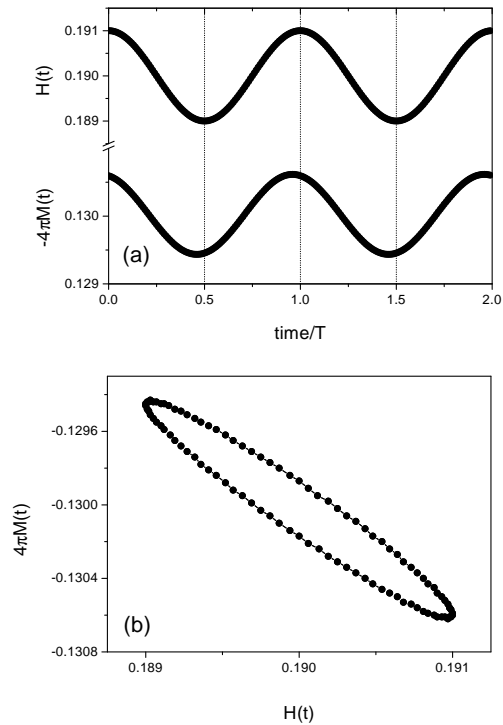}}
\caption{Response of a superconductor to a high frequency magnetic field.
(a) Time variation of the driving field and the sample magnetization and (b) the
magnetization loop that shows the presence of dissipation inside the sample.
(Sample size: $20\lambda\times20\lambda$, $H_{dc}=0.38H_{c2}(T)$,
$\omega t_0=0.028$).} 
\label{fig1}
\end{figure}\noindent

 We start by showing in 
Fig. 1 a typical high frequency ac magnetic response of a small sample. 
In Fig. 1(a) we plot the driving field $H(t)=H_{dc}+h_{ac}\cos\omega t$
 and the corresponding magnetization $M(t)$
of the sample. 
We show the response of the system after a long equilibration
time (typically around 100 periods of the external ac field).
It can be observed that the magnetization has a sinusoidal
behavior, which is a signal that we are in the linear regime and also 
that the sample is following the external perturbation. 
It can also be observed that there is a 
phase shift between the magnetization and the external field due to the 
presence of dissipation. 
Fig. 1(b) shows the magnetization loop obtained from the signals of Fig. 1(a). 
The area inside the loop is proportional to the time average of the energy 
dissipated in the sample and it is proportional to the imaginary part of 
the ac susceptibility $\chi''$. 
The appearance of dissipation 
changes the relationship between $M_{ac}(t)$ and $H_{ac}(t)$ and 
therefore the real part of the ac susceptibility,
$\chi '(H_{dc},\omega )$, 
will depend on  the frequency of
the ac signal as well as on the bias field $H_{dc}$.

\subsection{Frequency dependence}

In order to understand the different dynamical regimes
of the ac response, we study  first the dependence
of  $\tilde \chi=\chi '+i\chi"$ with frequency $\omega$ for different
values of the dc magnetic field $H_{dc}$. 
In Fig. 2 and Fig. 3  we show the imaginary and the real part of
the susceptibility, respectively. These curves
were obtained for a sample of $20 \lambda \times 20\lambda$.

Fig. 2(a) shows the low frequency behavior of $\chi''$ for several
values of the magnetic field.
Fig. 2(b) shows the full $\chi''(\omega)$ curves obtained  
for fields such that there are no vortices within the sample, {\it i.e.}
in  the Meissner state.
For fields above the first penetration field $H_p$, vortices enter
into the sample.
The full frequency dependence of $\chi "(\omega)$ in 
the presence of vortices, {\it i.e.} for fields $H_{dc}>H_p$, 
is shown in Fig. 2(c).

The simplest case to understand is the behavior 
when there is no external field applied, $H_{dc}=0$.
In this case the complex susceptibility agrees very well,
both qualitatively and quantitatively, with the 
two fluid model described in Sec. IIB.
For low frequencies, $\chi''$ has a linear
dependence with $\omega$ as given by Eq.~(\ref{Msmall}).
There is a maximum in $\chi''(\omega,0)$ at 
$\omega = \nu_o=1/t_0$. At high frequencies $\omega \gg \nu_0$
the dissipation  is due to the normal 
electrons and follows the expected dependence $\chi''\sim \omega^{-1/2}$
as given by Eq.~(\ref{Mlarge}), corresponding to
 the normal state skin depth effect.

For small magnetic fields within the Meissner state, $0<H_{dc}\ll H_p$,
the behavior is qualitatively similar to the $H_{dc}=0$ case,
with a peak at $\omega=\nu_0$. The main difference is that
the slope of the low $\omega$ linear dependence 
increases with magnetic field,
as can be seen in Fig.2(a). 
\begin{figure}
\centerline{\includegraphics[width=8.5cm]{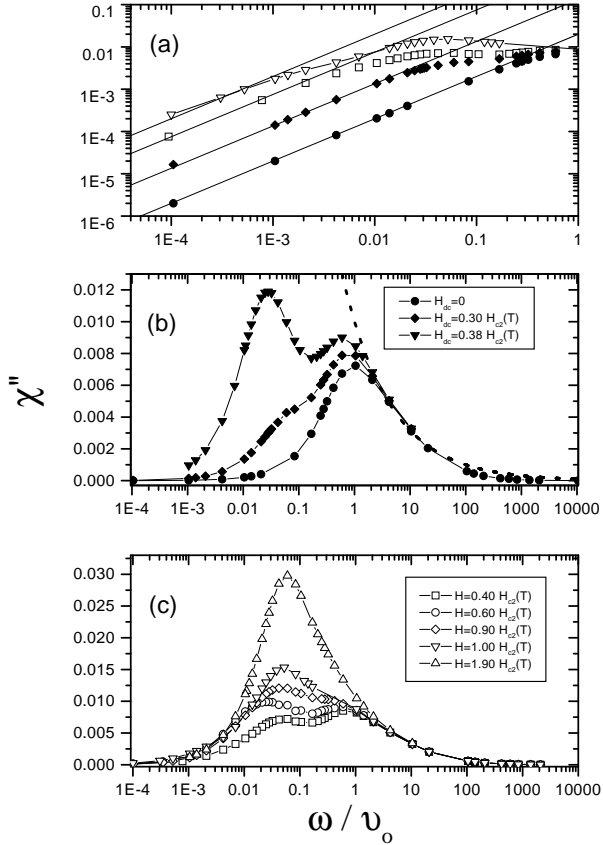}}
\caption{Frequency dependence of the imaginary component of the ac
susceptibility $\chi''$
at different fields $H_{dc}$, for samples of size $20\lambda\times20\lambda$.
(a) Low frequency behavior for fields in the Meissner state
(black symbols, $H_{dc}/H_{c2}(T)=0, 0.30$) and in the mixed state
(open symbols, $H_{dc}/H_{c2}(T)=0.40, 1.0$). The continuous 
lines correspond to a linear dependence $\chi''\sim \omega$.
(b) Frequency dependence in the Meissner state (number of vortices: $N_v=0$)
The dashed line corresponds to a dependence $\chi''\sim \omega^{-1/2}$.
(c) Frequency dependence in the mixed state  ($N_v\not=0$).
The frequency is normalized by $\nu_{o} = 1/t_0$. } 
\label{fig2} 
\end{figure}\noindent
According to Eq.~(\ref{Msmall}), 
the linear slope for low frequencies is proportional to $\lambda_L$. 
This implies that  the  London penetration depth should be
dependent on magnetic field: $\lambda_L(H)$. 
Also, Fig.2(b) shows that the magnitude of $\chi "(\omega)$
at any given $\omega$ increases for increasing $H_{dc}$.
These two effects are easily understood since
the Meissner screening currents
deplete the magnitude of the order parameter at the
surface, which leads
to a larger  static penetration depth $\lambda_L(H)$ 
(see also Sec.IVB) and 
to an increase in the normal electron dissipation.
The most interesting result is that for $H_{dc}$ near but still 
{\it below} $H_p$ a second dissipative maximum appears in $\chi "(\omega)$. 
This second 
peak is at a frequency $\omega_p$ two orders of magnitude below $\nu_o$. 

\begin{figure}[tb]
\centerline{\includegraphics[width=8.5cm]{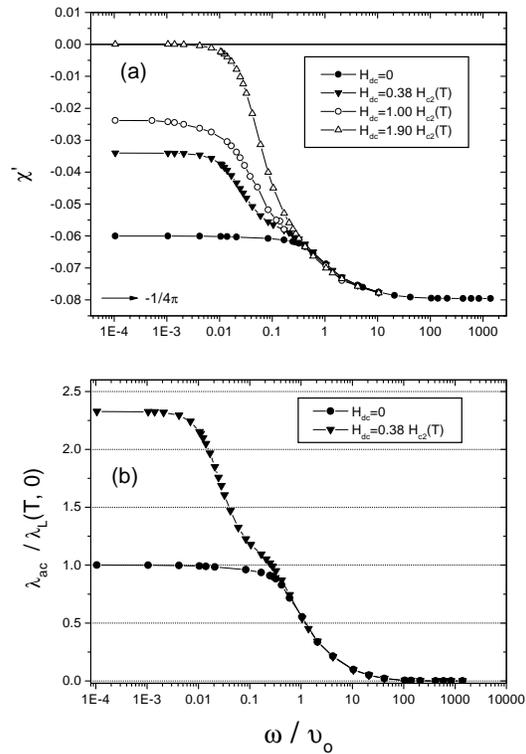}}
\caption{(a) Frequency dependence of the real component
 of the ac susceptibility $\chi'$ at different values of 
$H_{dc}$ (black symbols: Meissner state, open symbols: mixed state) and 
(b) effective ac penetration depth $\lambda_{ac}$ as a function of frequency.} 
\label{fig3} 
\end{figure}\noindent

For $H_{dc}>H_p$ 
the  vortices enter into the sample. 
In this case we still find a low frequency linear dependence 
$\chi'' \sim \omega$, as shown in Fig.2(a). 
Only for large fields close to $H_{c2}$ 
a small departure from linear dependence is observed, 
$\chi'' \sim \omega^\alpha$, with $\alpha\approx0.8$.
The behavior for very large frequencies $\omega\gg\nu_0$ 
is the same for all magnetic
fields, since it corresponds to the normal state skin depth
effect with $\chi'' \sim \omega^{-1/2}$, as mentioned before.
As can be observed in Fig.2(c), the main characteristic of 
the mixed state is that there are
two dissipation  maxima as a function of 
frequency in $\chi''(\omega)$. The high frequency maximum 
corresponds to the ``two fluid peak'' at $\omega\approx\nu_0$. 
The low frequency maximum that appears at a frequency $\omega_p \ll
\nu_0$, is a continuation of the maximum observed when $H_{dc}\alt
H_p$. The peak frequency $\omega_p$ shows a weak dependence with magnetic
field, $\omega_p$ first decreaseas and then increases with increasing
$H_{dc}$. For fields $H_{dc}\ge H_{c3}$ the sample is completely in the normal
state and there is a single maximum in the $\chi''$ vs $\omega$ curve.
It corresponds to the frequency at which the normal state skin deph $\delta_n$
equals the system size $L$.

Let us now analize the frequency dependence of
the real part of the susceptiibility, $\chi '(\omega)$, 
which is shown in Fig.~3(a) for some values of $H_{dc}$.
From $\chi '(\omega)$ it is possible to extract the real part
of the effective ac penetration depth $\lambda_{ac}=\lambda'(\omega)$
using Eq.~(\ref{musq}). The length $\lambda_{ac}$ represents
the length scale for the penetracion of the ac component of
the magnetic field and its frequency dependence is 
shown in Fig.~3(b).
In the absence of a bias field ($H_{dc}=0$), 
$\chi '(\omega)$ is well described by the two fluid model, as given
in Eqs.~(\ref{Mlarge}),(\ref{Msmall}).
The sample changes from a London screening for $\omega \ll \nu_{o}$,
with $\lambda_{ac}\approx\lambda_L$,
to a perfect diamagnetic state for $\omega \gg \nu_{o}$, with
$\lambda_{ac}\approx 0$. 
At low frequencies $\chi '$ is larger than  
$-1/4 \pi$ due to the penetration of the field in a region of size 
$\lambda_{L}$ near the sample surface,
$\chi'\approx(-1/4\pi)(1-4\lambda_L/L)$.
 At high frequency the ac penetration
depth is $\lambda_{ac}\ll\lambda_{L}$ and 
$\chi '(\omega)\approx-1/4 \pi$. 
The opposite case is for $H_{dc}=1.9 H_{c2}(T)>H_{c3}$ with the sample
mostly in the normal state
(at this field the surface superconductivity disappears 
and there is superconductivity only at the corner of the
sample\cite{enomoto}). 
As can be expected, in this case the system behaves as a normal metal
and is dominated by the skin-depth effect with
$\lambda_{ac}\approx\delta_n/2$. For low frequencies,
the skin-depth $\delta_n$ equals the system size and thus 
$\chi'\approx 0$.  
For magnetic fields within the Meissner state, $0<H_{dc}\ll H_p$,
we find that for low frequencies, $\omega\ll\nu_0$, the ac
penetration depth saturates to the field dependent London penetration
depth $\lambda_L(H)$, which increseas with $H$.
The most interesting case is for $H_{dc}\alt H_p$, which is shown
in Fig.~3. In this situation,
one can distinguish three frequency regimes, as it was found before 
in the dissipative part, $\chi''(\omega)$. At the frequencies
for which there are maxima in the dissipation, $\nu_0$ and $\omega_p$,
there is a rapid change in the value of $\chi'(\omega)$ (or in
$\lambda_{ac}(\omega)$). For $\omega\gg\nu_0$ the normal fluid behavior
of Eq.~(\ref{Mlarge}) is followed, as expected for all values of $H_{dc}$.
For the frequency range $\omega_p<\omega<\nu_0$, there is a shoulder in
$\chi'(\omega)$ (and $\lambda_{ac}({\omega})$), 
which basically corresponds to an ac penetration depth
of the order of the expected value for $\lambda_L(H)$ at that field.
At low frequencies, $\omega<\omega_p$, the ac penetration depth
saturates to a length $\ell_p$  which is smaller than the
system size and larger than $\lambda_L(H)$.

In summary, the most interesting feature observed in $\tilde\chi(\omega)$ 
is the appearance of a dissipation
maximum at frequencies $\omega_p(H)\ll\nu_0$ 
in the presence of a magnetic field.  A
dissipation maximum in macroscopic samples is expected when $H\not=0$,
as we discussed in Sec.~II.
It can be caused by several reasons: 
(i) finite size, (ii) surface barriers or (iii) bulk pinning.
However, the behavior predicted in any of these
cases is not followed in the mesoscopic samples studied.
Let us discuss each of the possibilities.
(i) Finite size: in a small sample this is the first effect to analyze.
The characteristic frequency for finite size effects, $\omega_L$,
should increase linearly with $B$ and decrease with system size, as
given by Eq.~(\ref{peakL}). However,   $\omega_p$ has a weak and
non-monotonous dependence with magnetic field, as seen in Fig.~2(c).
In Fig.~4  we show $\chi''(\omega)$ in a smaller sample  with
$L=5\lambda$ (and  $\eta=12$) for $H_{dc}\alt H_p$. We see that the maximum 
appears at the same frequency $\omega_p$  as in the $L=20\lambda$ case
for the same magnetic field [Fig.~2(b)].
\begin{figure}[tb]
\centerline{\includegraphics[width=8.5cm]{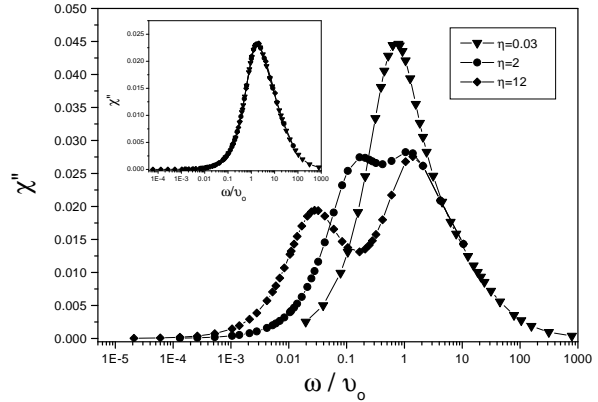}}
\caption{
Dependence with the TDGL time scale 
$ t_{GL} = \eta t_0$.
$\chi''$ vs. $\omega$ curves  for small mesoscopic samples
$5\lambda\times 5\lambda$ in the Meissner state ($N_v=0$) 
near the penetration field, $H\alt H_p$, $H_{dc}=0.35H_{c2}(T)$.
Inset shows the  dependence with $\eta$ for $H_{dc}=0$.
}
\label{fig4} 
\end{figure}\noindent
Therefore $\omega_p$ has no size dependence. 
Furthermore, in Fig.~3 we see that when $\omega\rightarrow0$ the sample
is still diamagnetic and $\lambda_{ac} < L$ (a finite size effect would
have $\chi'(\omega\rightarrow 0)\approx 0$ and $\lambda_{ac}\agt L$).
(ii) Surface barrier: in mesoscopic samples, surface barriers are
very important.\cite{HD} In macroscopic samples, the effect of
surface barriers, assuming a large number
of vortices, gives a low frequency dependence for dissipation as
$\chi''\sim\omega^{\alpha}$ with $\alpha=1/2$, 
see Eq.~(\ref{Ssmall}) and Ref.~\onlinecite{sonin}. However,
we find a linear frequency dependence in most cases as  shown in the
inset of Fig.~2(a), and even when there is a departure from linearity,
it is with an exponent of $\alpha\approx0.8>0.5$. Also the
characteristic frequency $\omega_{SB}$ of Eq.~(\ref{peakS}) has a 
strong dependence on magnetic field, which is not observed here.
(iii) Pinning: since there is no disorder in the model, no bulk pinning
is expected. However, in a very small sample one may argue
that, given the smallness of the system, the sample itself 
acts as a confinement potential for vortices. In this case, $\kappa_p$ in
Eq.~(\ref{vormot}) would represent the vortex elastic response for
small oscillations within this confinement potential. 
The characteristic frequency for pinning effects, $\omega_C$, is
independent of magnetic field in macroscopic samples, according to 
Eq.~(\ref{peakC}). The weak field dependence of $\omega_p$ may be
still consistent with this  result (the effective $\kappa_p$ may depend
on the number and distribution of vortices inside the sample 
in this hypothetic  case of ``confinement potential''). 
Also the fact that $\lambda_{ac} < L$ for 
$\omega\ll\omega_p$ in Fig.~3(b) is consistent with a finite
``Campbell's penetration depth'' $\lambda_C$ (and then
$\ell_p\equiv\lambda_C$).
However, we also have to discard this possibility, since 
we find that there is a dissipation
maximum at $\omega_p$ {\it even when there are no vortices in the
sample}.
As shown in Fig.~2(b) the dissipation peak already appears for
magnetic fields just below the penetration field $H_p$. To be
more precise,
we have explicitly calculated the topological number of
total vorticity as $N_v=\frac{1}{\Phi_o}\oint (A+\frac{J_s}{|\Psi|^2})dl$. 
We obtain $N_v=0$ at all times for the magnetic fields shown in Fig.~2(b).
Therefore, all arguments based on the oscillation of vortices, as for example
the Coffey and Clem model\cite{clem} of Eq.~(\ref{cofclem}), which work
very well for macroscopic samples, are not enough to explain the
dissipation maxima observed in these mesoscopic samples.

\begin{figure}[tb]
\centerline{\includegraphics[width=8.5cm]{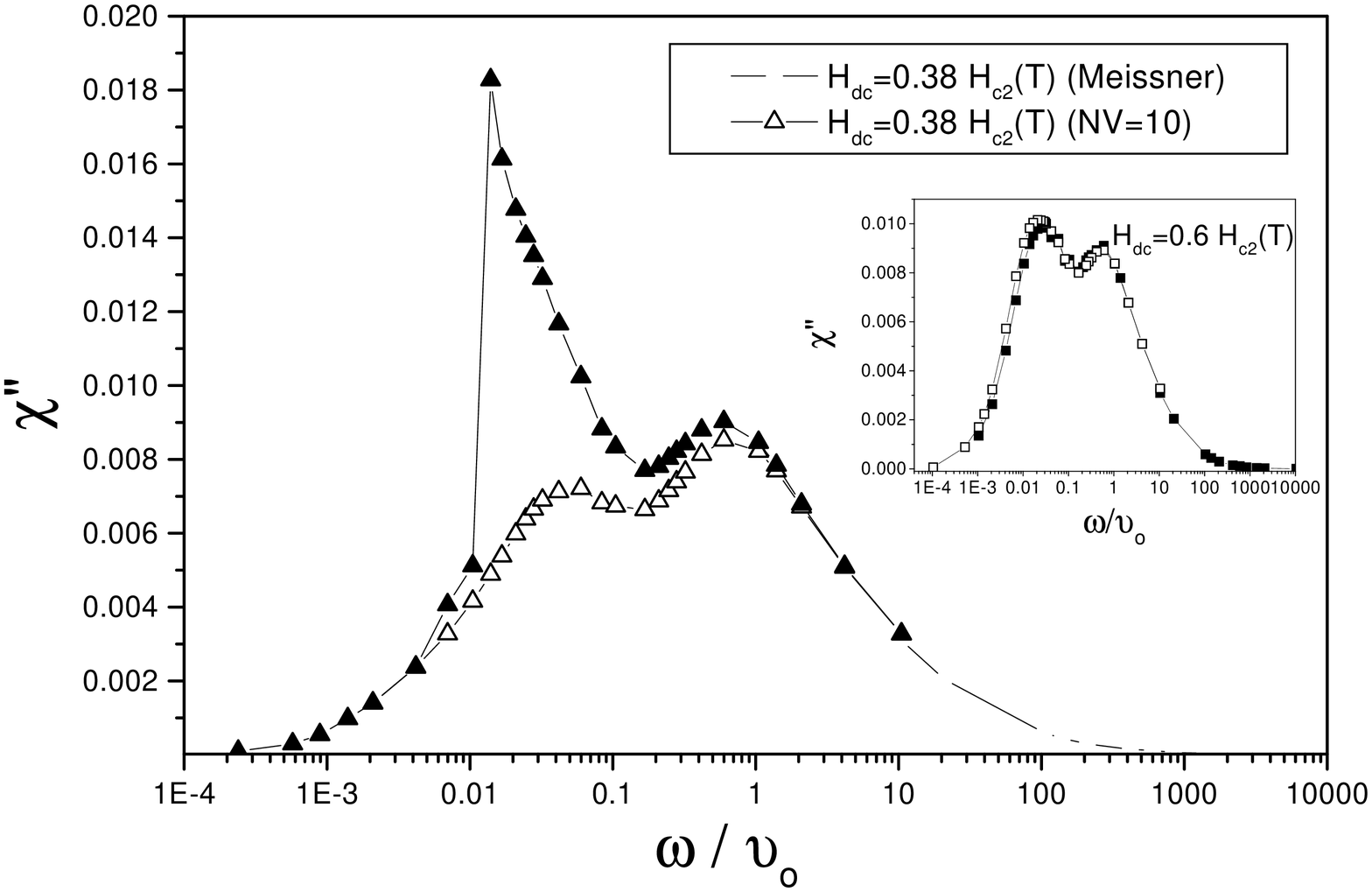}}
\caption{History dependence of $\chi''$ vs. $\omega$ curves 
for $H_{dc}$ near the penetration field $H_p$
($H_{dc}=0.38H_{c2}(T)$). Black triangles: decreasing
frequency from a high $\omega$ state without vortices ($N_v=0$);
the branch for $\omega$ above the jump has $N_v=0$ and below the 
jump $N_v=12$. Black symbols: increasing frequency from a low $\omega$
state with $N_v=12$. 
Sample size is $20\lambda\times20\lambda$.
The inset shows the absence of history  dependence for a
$H_{dc}=0.6H_{c2}(T)$.} 
\label{fig5} 
\end{figure}\noindent

In Fig.~4 we show that $\omega_p$ depends directly on the
time scale for the relaxation of the amplitude of the order parameter,
$t_{GL}=\eta t_0$.  
We consider a magnetic field $H_{dc}\alt H_p$
and we change the value of $\eta$ in Eqs.(\ref{tdgl1}) and (\ref{tdgl2}). 
The ``two fluid peak''
is always at the same frequency $\nu_0=1/t_0$, since $t_0$ is fixed.
We find that when decreasing $\eta$, the frequency $\omega_p$ shifts
to higher values and increases monotonically with $1/t_{GL}$,
until the rather unphysical case of $t_{GL}<t_0$ ($\eta<1$), when only
the two fluid peak at $\nu_0$ is observed. 
The dependence of $\omega_p$ with $t_{GL}$ shows 
that variations in the amplitude of the order parameter are
the main mechamism of this dissipation maximum, as we will see
more clearly later in Sec.~V.
In the inset of Fig.~4 we verify that for $H_{dc}=0$ the behavior
is independent of $t_{GL}$, since this case can be simply described 
with the two fluid model.

The appearance of the dissipation maximum for a magnetic field below $H_p$,
just before the penetration of vortices into the sample, leads to
hystory dependent effects. 
The curves in Fig.~2 were generated 
increasing the field $H_{dc}$ at low frequency and 
then measuring $\chi "(\omega)$ for increasing $\omega$. 
In this way, the curves reported in
Fig.~2(c) have the same  number of vortices at all frequencies
for a given $H_{dc}$.
If we follow the opposite procedure, {\it i.e.},
we increase $H_{dc}$ at high frequency and then measure $\chi "(\omega)$ 
for decreasing $\omega$, the curves can show discontinuous jumps at 
certain frequencies  values due to vortex penetration. 
This occurs because increasing the field 
at high frequency can generate a metaestable state  
that can become unstable at low frequencies.
In Fig.~5 we show this case, for a field near the first penetration
field $H_{dc}\approx H_p$. Depending on the history, the high frequency
branch can have either $N_v=12$ vortices or $N_v=0$. 
In the case when the frequency is decreased from a high frequency state,
there is a large jump in $\chi''(\omega)$ at a frequency near
$\omega_p$, 
from a metastable branch with $N_v=0$ for $\omega > \omega_p$  to
a branch with $N_v=12$ for $\omega < \omega_p$.
In contrast, we see in the inset of Fig.~5 that for a field 
$H_{dc}\not=H_p$ the curves of $\chi''(\omega)$ are independent of history.

\subsection{Magnetic field dependence}

Experimentally, the ac frequency is fixed and the dc magnetic
field or the temperature can be varied.
In this section we will study the ac magnetic response of mesoscopic 
superconductors varying the bias dc magnetic field. We will show 
results at different frequencies and for different sample sizes.

In the absence of the ac field,
the dc magnetic behavior of mesoscopic superconductors is different from the 
continuos macroscopic behavior.\cite{bolech,HD} 
In mesoscopic samples each vortex 
entrance event produces discontinuos jumps in the magnetization 
curve at certain $H_{dc}$ values. 
Each discontinuous jump in $M(H)$ corresponds
to a sudden increase in the number of vortices. 
These jumps occur at succesive magnetic fields 
$H_{pi} = H_{p1},  H_{p2}, H_{p3}, \ldots$.\cite{HD} 
In the regions of $H_{pi} < H < H_{p(i+1)}$ the
number of vortices $N_v$ is constant. The only penetration
events occur at $H_{pi}$, when vortices enters into the sample. 
In the region of constant vorticity,
$H_{pi} < H < H_{p(i+1)}$, one may think that no vortices enter
the sample because of the effect of the surface barrier.

Fig. 6 (a) shows the behavior of $\chi" (H_{dc})$ obtained at a fixed 
frequency $\omega= 0.004\nu_{o}<\omega_p$ in a mesoscopic sample of size 
$10\lambda \times 10\lambda$. 
At small $H_{dc}$, in the Meissner state, 
$\chi"$  increases continuously with increasing $H_{dc}$. 
The presence of the dc magnetic 
field in the Meissner state induces static supercurrents which deplete
the order parameter at the boundary. As explained before in Sec.~IVA, 
the depletion of the order parameter 
leads to an increase of dissipation for increasing $H_{dc}$. 
When the dc magnetic field is increased above the first
penetration field $H_{p}\equiv H_{p1}$, the 
first vortices enter into the sample. 
Fig. 6(a) shows that there is a discontinuous 
jump in $\chi"$ with a decrease of dissipation just at $H_{p}$.
Two states are possible exactly at $H_p$: one state without
vortices and high dissipation, and one sate with vortices inside
the sample and {\it low} dissipation. This is in agreement
with the history dependence observed in Fig.~5.
Further jumps in the dissipation curve $\chi" (H_{dc})$ are present at
the other magnetic fields for vortex penetration, 
$H_{p2},H_{p3},\dots$. The jumps  
are followed by a later continuos increase of $\chi" (H_{dc})$
with increasing $H_{dc}$ while the number of vortices $N_v$ remains fixed. 
Fig.~6(b) was obtained in a sample of the same size of 
Fig.~6(a) but at a higher frequency $\omega= 0.06 \nu_{o} > \omega_p$. 
\begin{figure}[tb]
\centerline{\includegraphics[width=8.5cm]{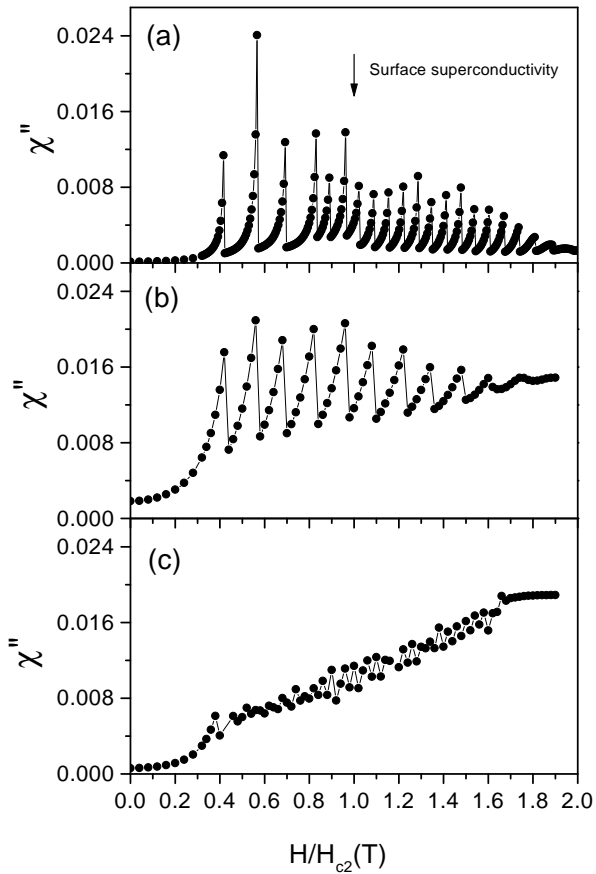}}
\caption{$H_{dc}$ dependence of $\chi "$ obtained at a fixed frequency.
Small mesoscopic samples of size $10 \lambda \times 10\lambda$ showing 
discontinuos jumps: 
(a) $\omega t_0=0.004$ and 
(b) $\omega t_0=0.06$.
(c) Large samples of size   $40 \lambda \times 40\lambda$
showing the  usual continuos behavior, $\omega t_0=0.06$.} 
\label{fig6} 
\end{figure}\noindent
The frequencies 
used in Figs.~6(a) and 6(b) are  at both sides of the 
dissipation maximum that appears at $\omega_p$. Both curves 
are similar in their qualitative features, but show a
few differences. At high magnetic fields there are less jumps in
$\chi''(H_{dc})$ in Fig.~6(b) than in Fig.~6(a).
This is because at high frequencies it is possible to remain in
a metastable state with a fixed number of vortices in a wider
range of magnetic field. 
At the same time, the higher frequency of Fig.~6(b) produces a 
decrease in the amplitude of the jumps. 
The behavior of $\chi''(H_{dc})$
of Figs.~6(a) and 6(b)  is different from the behavior of bulk samples. 
For example, 
Fig. 6(c) shows $\chi" (H_{dc})$ at $\omega= 0.06 \nu_{o}$ obtained in a 
``large'' sample of $40\lambda \times 40\lambda$. 
In this case there are no discontinuous jumps 
and the curve is almost continuous. 
In macroscopic samples the entrance of vortices for $H_{dc} > H_p$
increases the ac losses in a continuos way. 
In the macroscopic case vortices play a fundamental role
in the ac losses, and therefore 
an increase on the number of vortices increases the
dissipation. In contrast, the results of Figs.~6(a) and 6(b)
show the opposite behavior in mesoscopic samples: 
there is a sudden decrease of the ac losses at the magnetic fields
that correspond to an increase in the number of vortices.

\begin{figure}[tb]
\centerline{\includegraphics[width=8.5cm]{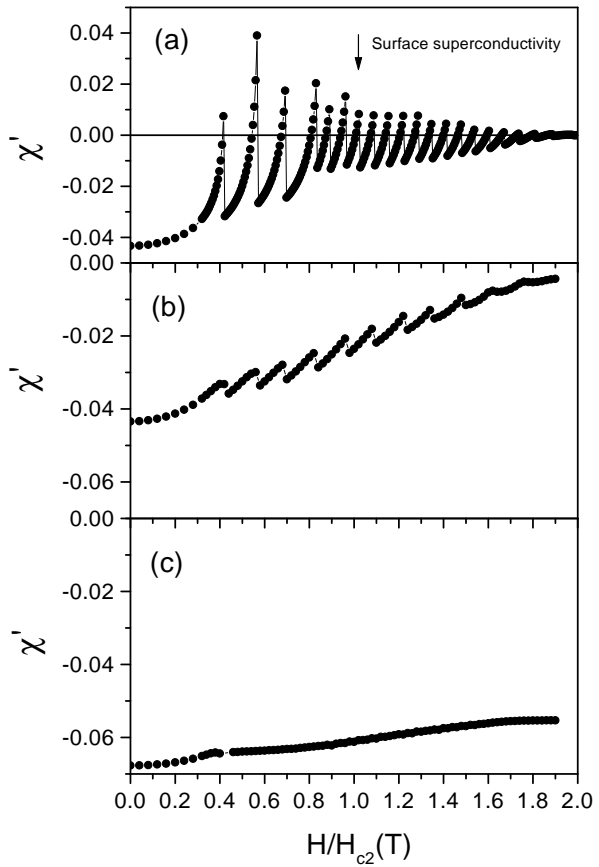}}
\caption{$H_{dc}$ dependence of $\chi '$ obtained at a fixed frequency:
(a) and (b) for small samples  and (c) for large samples. 
The parameters are the same as used in Fig.~6. } 
\label{fig7} 
\end{figure}\noindent

Fig.~7 shows the real part of the ac susceptibility for 
increasing $H_{dc}$. We use the same parameters as in Fig.~6. 
The curves of  $\chi'(H_{dc})$ also have a typical mesoscopic behavior with 
discontinuities at each vortex penetration field. The increase of frequency
from $\omega =0.004 \nu_o$ in Fig.~7(a) to $\omega =0.06 \nu_o$ in 
Fig. 7(b) shows similar differences  as observed in $\chi"(H_{dc})$. 
It can also be noted in Fig. 7(a) that the differential susceptibility has a 
paramagnetic behavior ($\chi' > 0$) near each vortex penetration field. 
This can be expected from the behavior of the dc magnetization curves of 
mesoscopic samples. \cite{bolech} 
In mesoscopic samples, for $H_{dc}$ below each 
vortex penetration field  a magnetization maximum appears and 
therefore a region near $H_{pi}$ where the magnetization increases at 
increasing field. 
This shows in the low frequency 
ac response of mesoscopic samples as a differential paramagnetic behavior. 
Fig. 7(c) shows a continuos macroscopic dependence of $\chi'(H_{dc})$ in 
a larger sample of size $40\lambda \times 40 \lambda$. 
At the frequencies used in Fig. 7(c) we can see a diamagnetic behavior  
even in the normal state for fields greater than $2 H_{c2}(T)$.
This is simply because 
at high frequencies there is magnetic screening in the normal state
due to the skin depth effect. 

\begin{figure}[tb]
\centerline{\includegraphics[width=8.5cm]{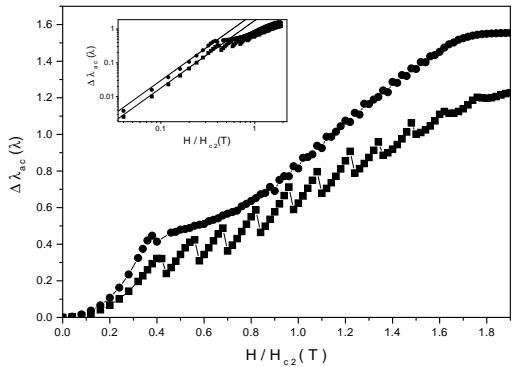}}
\caption{Change in the ac penetration depth ($\Delta\lambda_{ac}$) 
in a mesoscopic  sample (squares, $10 \lambda \times 10\lambda$) 
and a macroscopic sample (circles, $40 \lambda \times 40\lambda$)
for $\omega t_0=0.06$. 
 The inset shows a fit to a quadratic dependence
 $\Delta\lambda_{ac}\sim H^2$ (continuos lines)
 for fields below $H_p$.} 
\label{fig8} 
\end{figure}\noindent

From Fig.~7 it is possible to obtain the $H_{dc}$ dependence of the London 
penetration depth. This dependence has been a subject of interest 
recently because of its relation with the simmetry of the order parameter. 
In conventional s-wave superconductors the following dependence is
expected:
\begin{equation} 
\Delta \lambda (H_{dc},T) / \lambda (0,T) = \beta(T) (H_{dc}/H_{c2} (T))^2 
\label{lambdah}
\end{equation}
where $\Delta \lambda (H_{dc},T)= \lambda (H_{dc},T) - \lambda (0,T)$. It is
assumed that the sample is in the Meissner state and in the linear regime, 
i.e. $|H_{dc}| \gg |h_{ac}|$.
Eq.~(\ref{lambdah}) was originally obtained by Ginzburg and 
Landau \cite{landau} and also by Bardeen \cite{bardeen} an it 
has been observed 
experimentally in type I \cite{pippard,sridhar2} and conventional type II 
superconductors \cite{maeda1}. 
The behavior of unconventional superconductors 
is expected to be different \cite{maeda2}. In the case of a d-wave 
superconductor Yip and Sauls \cite{yip1,yip2} proposed 
a linear field dependence 
$\Delta \lambda (H_{dc}) \propto H_{dc}$.
Here we use the phenomenological TDGL equation in its s-wave form.
Therefore, in the 
Meissner state we expect a result similar to Eq.(\ref{lambdah}) for macroscopic
samples. The results are shown in 
Fig.~8. Both the macroscopic and the mesoscopic sample 
show a quadratic dependence  of $\Delta \lambda(H_{dc})$ as a function 
of $H_{dc}$ in the Meissner state. The vortex
penetration produces a sudden decrease of $\Delta\lambda$ in mesoscopic samples, 
whereas at a fixed number of vortices $\Delta \lambda$ increases continuously 
for increasing $H_{dc}$.  
For macroscopic samples the entrance of vortices at
the first penetration field $H_{p}$
does not produce discontinuity in $\Delta\lambda(H_{dc})$ but a change 
of slope as can be observed in the  plot. For $H_{dc}>H_{p}$ the 
dependence changes from quadratic to approximately linear in $H_{dc}$.

\section{NASCENT VORTICES EFFECT}

From the results of the previous section, it is clear now that
the ac magnetic response of a mesoscopic superconductor is different from the
macroscopic behavior and it can not be explained with the known models. 
An interesting result is the significant decrease of dissipation each time 
the number of vortices increases. 
To understand this effect we have analyzed in detail
the time variation of the order parameter and the
magnetization within the sample.

\begin{figure}[tb]
\centerline{\includegraphics[width=8.5cm]{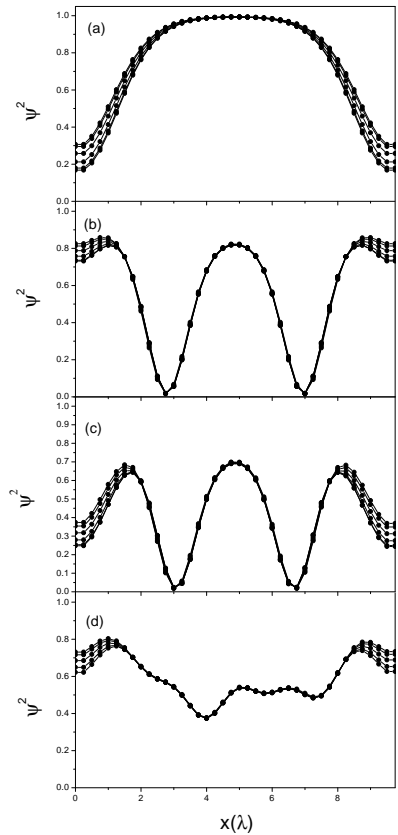}}
\caption{Time variation of the order parameter in a sample of size 
$10 \lambda \times 10 \lambda$ at different applied magnetic fields
and for $\omega t_0=0.06$.
 The figures show 
transversals cuts taken at the center of one of the faces for
different times. (a) $H_{dc}=0.4216H_{c2}(T)<H_{p1}$, 
(b) $H_{dc}=0.4217H_{c2}(T)>H_{p1}$, (c) $H_{dc}=0.568H_{c2}(T)<H_{p2}$, 
(d) $H_{dc}=0.569H_{c2}(T)>H_{p2}$.}
\label{fig9} 
\end{figure}\noindent

Fig. 9 shows the time variation of the order 
parameter at different applied magnetic fields in a small sample.
Each curve represents the amplitude of the order
parameter along the $x$ direction,
perpendicular to a face of the square sample, obtained at a given time. 
To simplify the analysis the cut was done just at 
the center of the face at $y=L/2$ 
(which is the preferred point for vortex
penetration in a square sample). 
The order parameter is obtained at different times 
within a single period of the ac magnetic field. 
In particular, we show the 
behavior of the order parameter just before and after the two 
first values of the penetration fields, $H_{p1}$ and $H_{p2}$. 
At $H_{dc}=0$ the order parameter $|\Psi|$ is equal to unity in 
the entire sample and the dissipation is small. 
In the Meissner state but for
$H_{dc} \alt H_{p1}$, Fig. 9 (a)  shows that the magnitude 
of the order parameter near the boundary, and at the center of
the face, has large oscillations around a small value. 
 Fig. 9 (b) shows the
time evolution of order parameter just after the first vortex entrance
event, $H_{dc} \agt H_{p1}$. 
We see that the entrance of vortices increases the order parameter at the
boundary, where now $|\Psi|$ oscillates around a value much higher
than before.
One can also see that the positions of the vortices are fixed
inside the sample and do not oscilllate in the presence of the ac signal. 
Recently, in an extension of the idea of the Bean
Livingston barrier to mesoscopic superconductos, we have shown that the 
vortex entrance produces a reinforcement of the surface barrier,
due to the interaction between the vortices that are trying to enter the sample
and the vortices already inside the sample.\cite{HD}
The presence of this interaction decreases the surface currents.
Thus, the main effect  after the entrance of vortices 
inside a mesoscopic superconductor is that the Meissner currents decrease
and the order parameter increases at the boundary. 
Therefore the normal state electrons decrease 
and then the dissipation decreases. 

Figs.~9(c)  and 9(d) show similar behavior for $H_{dc}\alt H_{p2}$
and $H_{dc}\agt H_{p2}$. After increasing 
$H_{dc}$ at a constant number of vortices,  the order parameter at the
boundary has decreased, as seen in Fig.~9(c). 
The decrease of the order parameter increases the dissipation as was 
observed before. If new
vortices enter again into the sample, 
the order parameter increases at the boundary, see Fig.~9(d) and
therefore the dissipation decreases again.
This process is repeated in the other penetration fields $H_{pi}$. 
On the contrary, the entrance 
of vortices in a large sample do not produce an appreciable change in the 
amplitude of the order parameter at the boundary 
before and after the penetration event, and therefore the dissipation 
shows a continuous behavior [as seen in Fig.~6(c)]. 

A careful analysis of the entire profile of the order parameter for magnetic 
fields below but near to $H_{p}$ suggest
that the depletion of  the order parameter near the boundary
may be attributed to ``vortices'' that are about to enter
into the sample. These ``nascent vortices'' are located outside of the 
sample but near the surface. 
Even more, for higher values of $h_{ac}$  (beyond linear response)
we find  that these ``nascent 
vortices''  oscillate near the surface
({\it i.e.} that the profile of the depletion of $|\Psi|$
near the surface oscillates with $H(t)$). 
According to this scenario, 
the length scale $\ell_p$ observed in Fig.~3(b) for low
frequencies may correspond to a characteristic length scale for
the nucleation of vortices from the boundary. (For large fields $B\gg
H_p$, the length $\ell_p$ could be related to 
the size of the vortex free region
$d_{SB}$ of macroscopic samples).
Years ago, Walton and Rosemblum\cite{walton} 
suggested the idea of nascent vortices as a possible source of 
the high-frequency losses in superconductors. This idea had further support 
from the analytical work of Kramer \cite{kramer} who found static solutions 
of the Ginzburg-Landau equations which corresponded to
vortices nucleated near the sample. 
The nascent vortices can oscillate under an ac field.
This source
of dissipation is not generally taken into account in macroscopic samples
because  in this case is small compared with the dissipation due to the motion 
of vortices inside the sample. 
However in a mesoscopic superconductor, 
with a few vortices confined inside the sample, 
this source of dissipation can be a significative part of the 
dissipation observed, as we have found here.

\section{SUMMARY AND CONCLUSION}

We have studied the response of mesoscopic superconductors to an 
ac magnetic field. 
In mesoscopic superconductors vortices are confined inside the sample 
by the surface barrier and their dynamics 
play a secondary role in the dissipation. 
When increasing $H_{dc}$, discontinuities in $\chi'$ and $\chi"$ 
appear at each vortex penetration event. The dissipation
is maximum before the penetration of vortices due to the effect
of ``nascent vortices'', which lead to large oscillations of the
amplitude of the order parameter at the boundary.
After the penetration of vortices, the normal currents decreases and 
therefore the dissipation decreases, 
leading to a sudden suppression of the ac losses induced by 
the vortex entrance.

\begin{acknowledgments}

A.D.H. acknowledges Oscar Ar\'{e}s and C. Hart for useful
comments and help and the Centro Latino-Americano de F\'{\i}sica 
(CLAF) for financial support.  D.D. acknowledges
support from Conicet and CNEA.
We acknowledge
the financial support  from the Argentina-Cuba 
cooperation project SETCIP-CITMA: CU/A00-EIII/005 and from ANPCyT
(Argentina) project number PICT99-03-06343.

\end{acknowledgments}


\end{document}